\documentclass[twocolumn,pra,amsmath,amssymb]{revtex4}
\usepackage{graphicx}
\usepackage{dcolumn}
\usepackage{bm}
\newcolumntype{.}{D{.}{.}{-1}}

\begin{document}

\preprint{APS/123-QED}

\title{Nuclear Spin Effects in Optical Lattice Clocks}

\author{Martin M. Boyd, Tanya Zelevinsky, Andrew D. Ludlow, Sebastian Blatt, Thomas Zanon-Willette, Seth M. Foreman, and Jun Ye}
 \affiliation{JILA, National Institute of Standards and Technology and University of Colorado,
  Department of Physics, University of Colorado, Boulder, CO 80309-0440}

\date{\today}

\begin{abstract}
We present a detailed experimental and theoretical study of the
effect of nuclear spin on the performance of optical lattice clocks.
With a state-mixing theory including spin-orbit and hyperfine
interactions, we describe the origin of the $^1S_0$-$^3P_0$ clock
transition and the differential $g$-factor between the two clock
states for alkaline-earth(-like) atoms, using $^{87}$Sr as an
example.  Clock frequency shifts due to magnetic and optical fields
are discussed with an emphasis on those relating to nuclear
structure.  An experimental determination of the differential
$g$-factor in $^{87}$Sr is performed and is in good agreement with
theory.  The magnitude of the tensor light shift on the clock states
is also explored experimentally. State specific measurements with
controlled nuclear spin polarization are discussed as a method to
reduce the nuclear spin-related systematic effects to below
10$^{-17}$ in lattice clocks.
\end{abstract}

\maketitle

Optical clocks \cite{Diddams1} based on alkaline-earth atoms
confined in an optical lattice \cite{KatoriNature} are being
intensively explored as a route to improve state of the art clock
accuracy and precision. Pursuit of such clocks is motivated mainly
by the benefits of Lamb-Dicke confinement which allows high spectral
resolution \cite{Boyd1, HoytEFTF}, and high accuracy
\cite{Ludlow1,Boyd2,LeTargat1,KatoriJSP} with the suppression of
motional effects, while the impact of the lattice potential can be
eliminated using the Stark cancelation technique
\cite{Katori1,Katori2,Fortson1, Anders1}.  Lattice clocks have the
potential to reach the impressive accuracy level of trapped ion
systems, such as the Hg$^{+}$ optical clock \cite{Bergquist}, while
having an improved stability due to the large number of atoms
involved in the measurement. Most of the work performed thus far for
lattice clocks has been focused on the nuclear-spin induced
$^1S_0$-$^3P_0$ transition in $^{87}$Sr.  Recent experimental
results are promising for development of lattice clocks as high
performance optical frequency standards.  These include the
confirmation that hyperpolarizability effects will not limit the
clock accuracy at the $10^{-17}$ level \cite{Anders1}, observation
of transition resonances as narrow as 1.8 Hz \cite{Boyd1}, and the
excellent agreement between high accuracy frequency measurements
performed by three independent laboratories
\cite{Ludlow1,Boyd2,LeTargat1, KatoriJSP} with clock systematics
associated with the lattice technique now controlled below
10$^{-15}$ \cite{Boyd2}.  A main effort of the recent accuracy
evaluations has been to minimize the effect that nuclear spin
($I=9/2$ for $^{87}$Sr) has on the performance of the clock.
Specifically, a linear Zeeman shift is present due to the same
hyperfine interaction which provides the clock transition, and
magnetic sublevel-dependent light shifts exist, which can complicate
the stark cancelation techniques. To reach accuracy levels below
$10^{-17}$, these effects need to be characterized and controlled.

The long coherence time of the clock states in alkaline earth atoms
also makes the lattice clock an intriguing system for quantum
information processing. The closed electronic shell should allow
independent control of electronic and nuclear angular momenta, as
well as protection of the nuclear spin from environmental
perturbation, providing a robust system for coherent
manipulation\cite{Lukin06}. Recently, protocols have been presented
for entangling nuclear spins in these systems using cold collisions
\cite{Deutsch1} and performing coherent nuclear spin operations
while cooling the system via the electronic transition
\cite{Deutsch2}.

Precise characterization of the effects of electronic and nuclear
angular-momentum-interactions and the resultant state mixing is
essential to lattice clocks and potential quantum information
experiments, and therefore is the central focus of this work. The
organization of this paper is as follows. First, state mixing is
discussed in terms of the origin of the clock transition as well as
a basis for evaluating external field sensitivities on the clock
transition. In the next two sections, nuclear-spin related shifts of
the clock states due to both magnetic fields and the lattice
trapping potential are discussed. The theoretical development is
presented for a general alkaline-earth type structure, using
$^{87}$Sr only as an example (Fig.~\ref{structure}), so that the
results can be applied to other species with similar level
structure, such as Mg, Ca, Yb, Hg, Zn, Cd, Al$^+$, and In$^+$.
Following the theoretical discussion is a detailed experimental
investigation of these nuclear spin related effects in $^{87}$Sr,
and a comparison to the theory sections. Finally, the results are
discussed in the context of the performance of optical lattice
clocks, including a comparison with recent proposals to induce the
clock transition using external fields in order to eliminate nuclear
spin effects \cite{Hong,Santra,NistYb,NISTYb2,Thom,Katorimix}. The
appendix contains additional details on the state mixing and
magnetic sensitivity calculations.
\section{State Mixing in the $nsnp$ Configuration}
To describe the two-electron system in intermediate coupling, we
follow the method of Breit and Wills \cite{Breit} and Lurio
\cite{Lurio1} and write the four real states of the $ns~np$
configuration as expansions of pure spin-orbit (LS) coupling states,
\begin{equation}
{\small \begin{aligned}
|^3P_0\rangle &= |^3P_0^{0}\rangle\\
|^3P_1\rangle &= \alpha|^3P_1^{0}\rangle + \beta|^1P_1^{0}\rangle\\
|^3P_2\rangle &= |^3P_2^{0}\rangle
\\ |^1P_1\rangle &= -\beta|^3P_1^{0}\rangle +
\alpha|^1P_1^{0}\rangle.\\
\end{aligned}}\label{LSstate}
\end{equation}
\begin{figure}[t]
\resizebox{8.5cm}{!}{
\includegraphics[angle=0]{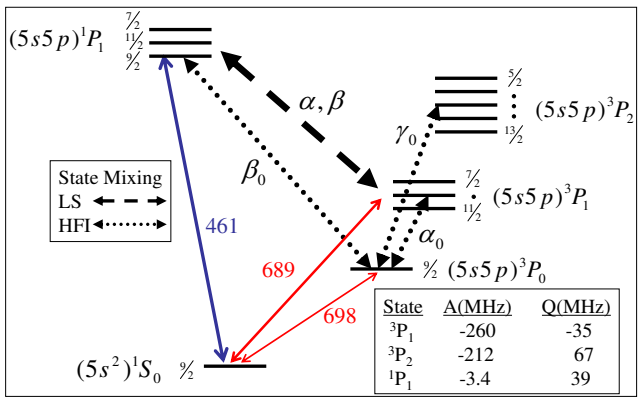}}
\caption{\label{structure}(color online) Simplified $^{87}$Sr energy
level diagram (not to scale).  Relevant optical transitions
discussed in the text are shown as solid arrows, with corresponding
wavelengths given in nanometers.  Hyperfine structure sublevels are
labeled by total angular momentum $F$, and the magnetic dipole ($A$)
and electric quadrupole ($Q$, equivalent to the hyperfine $B$
coefficient) coupling constants are listed in the inset. State
mixing of the $^1P_1$ and $^3P_1$ states due to the spin-orbit
interaction is shown as a dashed arrow. Dotted arrows represent the
hyperfine induced state mixing of the $^3P_0$ state with the other
$F=9/2$ states in the $5s5p$ manifold. }
\end{figure}

\noindent Here the intermediate coupling coefficients $\alpha$ and
$\beta$ (0.9996 and -0.0286 respectively for Sr) represent the
strength of the spin-orbit induced state mixing between singlet and
triplet levels, and can be determined from experimentally measured
lifetimes of $^1P_1$ and $^3P_1$ (see Eq.~\ref{alphabeta} in the
appendix). This mixing process results in a weakly allowed
$^1S_0$-$^3P_1$ transition (which would otherwise be
spin-forbidden), and has
 been used for a variety of experiments spanning different fields of atomic physics.
In recent years, these intercombination transitions have provided a
unique testing ground for studies of narrow-line cooling in Sr
\cite{Vogel,Mukaiyama1, Loftus2, Loftus1, Tino} and Ca
\cite{oatescool, sterrcool}, as well as the previously unexplored
regime of photoassociation using long lived states \cite{Zelevinsky,
YbPAS, JuliennePAS}.  These transitions have also received
considerable attention as potential optical frequency standards
\cite{OatesNew, SterrNew, Ido1}, owing mainly to the high line
quality factors and insensitivity to external fields. Fundamental
symmetry measurements, relevant to searches of physics beyond the
standard model, have also made use of this transition in Hg
\cite{Romalis}.  Furthermore, the lack of hyperfine structure in the
bosonic isotopes ($I=0$) can simplify comparison between experiment
and theory.

The hyperfine interaction (HFI) in fermionic isotopes provides an
additional state mixing mechanism between states having the same
total spin $F$,  mixing the pure $^3P_0$ state with the $^3P_1$,
$^3P_2$ and $^1P_1$ states.
\begin{equation}
{\small |^3P_0\rangle=  |^3P_0^{0}\rangle + \alpha_0|^3P_1\rangle
+\beta_0|^1P_1\rangle +\gamma_0|^3P_2^{0}\rangle}.\label{3p0state}
\end{equation}
The HFI mixing coefficients $\alpha_0$, $\beta_0$, and $\gamma_0$
($2\times10^{-4}$, $-4\times10^{-6}$, and $4\times10^{-6}$
respectively for $^{87}$Sr) are defined in Eq.~\ref{HFIdef} of the
appendix and can be related to the hyperfine splitting in the $P$
states, the fine structure splitting in the $^3P$ states, and the
coupling coefficients $\alpha$ and $\beta$ \cite{Breit, Lurio1}. The
$^3P_0$ state can also be written as a combination of pure states
using Eq.~\ref{LSstate},
\begin{equation}
{\small\begin{aligned}
|^3P_0\rangle =  &|^3P_0^{0}\rangle + (\alpha_0\alpha-\beta_0\beta)|^3P_1^{0}\rangle\\
& +
(\alpha_0\beta+\beta_0\alpha)|^1P_1^{0}\rangle+\gamma_0|^3P_2^{0}\rangle.
\end{aligned}}\label{3p0LSstate}
\end{equation}
The HFI mixing enables a non-zero electric-dipole transition via the
pure $^1P_1^{0}$ state, with a lifetime which can be calculated
given the spin-orbit and HFI mixing coefficients, the $^3P_1$
lifetime, and the wavelengths ($\lambda$) of the $^3P_0$ and
 $^3P_1$ transitions from the ground state \cite{Becker}.
\begin{equation}
{\small\tau^{^3P_0}=\left(\frac{\lambda^{^3P_0-^1S_0}}{\lambda^{^3P_1-^1S_0}}
\right)^3\frac{\beta^{2}}{(\alpha_0\beta+\beta_0\alpha)^2}\tau^{^3P_1}}.\label{3p0lifetime}
\end{equation}
In the case of Sr, the result is a natural lifetime on the order of
100 seconds \cite{Katori1, Porsev, Greene}, compared to that of a
bosonic isotope where the lifetime approaches 1000 years
\cite{Greene}.  Although the ~100 second coherence time of the
excited state exceeds other practical limitations in current
experiments, such as laser stability or lattice lifetime, coherence
times approaching one second have been achieved \cite{Boyd1}. The
high spectral resolution has allowed a study of nuclear-spin related
effects in the lattice clock system discussed below.

The level structure and state mixing discussed here are summarized
in a simplified energy diagram, shown in Fig.~\ref{structure}, which
gives the relevant atomic structure and optical transitions for the
$5s5p$ configuration in $^{87}$Sr.
\section{The Effect of External Magnetic Fields }
With the obvious advantages in spectroscopic precision of the
$^1S_0$-$^3P_0$ transition in an optical lattice, the sensitivity of
the clock transition to external field shifts is a central issue in
developing the lattice clock as an atomic frequency standard. To
evaluate the magnetic sensitivity of the clock states, we follow the
treatment of Ref. \cite{Lurio1} for the intermediate coupling regime
described by Eqns.~\ref{LSstate}-\ref{3p0LSstate} in the presence of
a weak magnetic field. A more general treatment for the case of
intermediate fields is provided in the appendix. The Hamiltonian for
the Zeeman interaction in the presence of a weak magnetic field $B$
along the $z$-axis is given as
\begin{equation}
{\small H_Z = (g_s S_z +g_l L_z - g_I I_z)\mu_0B.}\label{Hzeeman}
\end{equation}
Here $g_s\simeq 2$ and $g_l=1$ are the spin and orbital angular
momentum $g$-factors, and $S_z$, $L_z$, and $I_z$ are the
$z$-components of the electron spin, orbital, and nuclear spin
angular momentum respectively. The nuclear $g$-factor, $g_I$, is
given by {\small$g_I$=$\frac{\mu_I(1-\sigma_d)}{\mu_0 |I|}$}, where
$\mu_I$ is the nuclear magnetic moment, $\sigma_d$ is the
diamagnetic correction and {\small$\mu_0$=$\frac{\mu_B}{h}$}.
 Here, $\mu_B$ is the Bohr magneton, and $h$ is Planck's constant.  For
$^{87}$Sr, the nuclear magnetic momement and diamagnetic correction
are $\mu_I=-1.0924(7)\mu_N$ \cite{Olshewski} and $\sigma_d=0.00345$
\cite{Kopfermann} respectively, where $\mu_N$ is the nuclear
magneton. In the absence of state mixing, the $^3P_0$ $g$-factor
would be identical to the $^1S_0$ $g$-factor (assuming the
diamagnetic effect differs by a negligible amount for different
electronic states), equal to $g_I$. However since the HFI modifies
the $^3P_0$ wavefunction, a differential $g$-factor, $\delta g$,
exists between the two states. This can be interpreted as a
paramagnetic shift arising due to the distortion of the electronic
orbitals in the triplet state, and hence the magnetic moment
\cite{Lahaye}. $\delta g$ is given by
\begin{equation}
{\small\begin{aligned}
\\ \delta g  &= - \frac{\langle ^3P_0 |H_Z|^3P_0\rangle-\langle ^3P_0^{0}
|H_Z|^3P_0^{0}\rangle }{m_F \mu_0 B}
\\ =& - 2\left(\alpha_0\alpha-\beta_0\beta\right)\frac{\langle ^3P_0^{0}, m_F|H_Z|^3P_1^{0}, F=I,
m_F\rangle}{m_F \mu_0 B}
\\&+ \mathcal{O}(\alpha_0^2, \beta_0^2, \gamma_0^2, \ldots).
\end{aligned}}\label{dgeq}
\end{equation}
Using the matrix element given in the appendix for $^{87}$Sr
($I=9/2$), we find {\small $ \langle ^3P_0^{0}, m_F|H_Z|^3P_1^{0},
F=\frac{9}{2}, m_F\rangle$= $ \frac{2}{3}\sqrt{\frac{2}{33}}m_F
\mu_0 B$}, corresponding to a modification of the $^3P_0$ $g$-factor
by $\sim$60\%.  Note that the sign in Eq.~\ref{dgeq} differs from
that reported in \cite{Lahaye, Becker} due to our choice of sign for
the nuclear term in the Zeeman Hamiltonian (opposite of that found
in Ref. \cite{Lurio1}). The resulting linear Zeeman shift
{\small$\Delta_B^{(1)}$= $-\delta g m_F \mu_0 B$} of the
$^1S_0$-$^3P_0$ transition is on the order of
{\small$\sim$110$\times m_F$ Hz/G} (1 G = $10^{-4}$ Tesla). This is
an important effect for the development of lattice clocks, as stray
magnetic fields can broaden the clock transition (deteriorate the
stability) if multiple sublevels are used. Furthermore, imbalanced
population among the sublevels or mixed probe polarizations can
cause frequency errors due to line shape asymmetries or shifts. It
has been demonstrated that if a narrow resonance is achieved ($10$
Hz in the case of Ref. \cite{Boyd2}), these systematics can be
controlled at 5$\times10^{-16}$ for stray fields of less than 5 mG.
To reduce this effect, one could employ narrower resonances or
magnetic shielding.

An alternative measurement scheme is to measure the average
transition frequency between $m_F$ and $-m_F$ states of  to cancel
the frequency shifts.  This requires application of a bias field to
resolve the sublevels, and therefore the second order Zeeman shift
{\small$\Delta_B^{(2)}$} must be considered.  The two clock states
are both $J=0$ so the shift {\small$\Delta_B^{(2)}$} arises from
levels separated in energy by the fine-structure splitting, as
opposed to the more traditional case of alkali(-like) atoms where
the second order shift arises from nearby hyperfine levels.  The
shift of the clock transition is dominated by the interaction of the
$^3P_0$ and $^3P_1$ states since the ground state is separated from
all other energy levels by optical frequencies. Therefore, the total
Zeeman shift of the clock transition {\small$\Delta_B$} is given by
\begin{equation}
{\small\begin{aligned}
  \Delta_B & = \Delta_B^{(1)} +\Delta_B^{(2)} \\
 &= \Delta_B^{(1)} -\sum_{F'} \frac{|\langle ^3P_0, F,
m_F|H_Z|^3P_1,
F', m_F\rangle|^2}{\nu_{^3P_1, F'}-\nu_{^3P_0}}.\\
\end{aligned}}\label{Zeemanshift}
\end{equation}
The frequency difference in the denominator is mainly due to the
fine-structure splitting and is nearly independent of $F'$, and can
therefore be pulled out of the summation.  In terms of the pure
states, and ignoring terms of order $\alpha_0$, $\beta_0$,
$\beta^2$, and smaller, we have
\begin{equation}
{\small\begin{aligned}
  \Delta_B^{(2)} \simeq &  -\alpha^2 \frac{ \sum_{F'}|\langle ^3P_0^{0}, F, m_F|H_Z|^3P_1^{0},
 F', m_F\rangle|^2}{\nu_{^3P_1}-\nu_{^3P_0}}\\
 &=  -\frac{2\alpha^2(g_l-g_s)^2\mu_0^2}{3(\nu_{^3P_1}-\nu_{^3P_0})}B^2,\\
\end{aligned}}\label{secondorder}
\end{equation}
\begin{figure}[t]
\resizebox{8.5cm}{!}{
\includegraphics[angle=0]{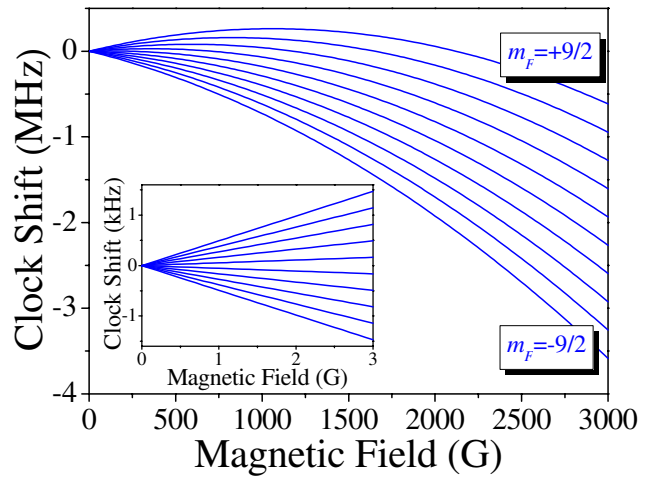}}
\caption{\label{3p0zeeman}(color online) A Breit-Rabi diagram for
the $^1S_0$-$^3P_0$ clock transition using Eq.~\ref{BRLike-formulae}
with $\delta g \mu_0=-109$ Hz/G. Inset shows the linear nature of
the clock shifts at the fields relevant for the measurement
described in the text.}
\end{figure}
\noindent where we have used the matrix elements given in the
appendix for the case $F=9/2$.  From Eq.~\ref{secondorder} the
second order Zeeman shift (given in Hz for a magnetic field given in
Gauss) for $^{87}$Sr is {\small$\Delta_B^{(2)}$=$-0.233 B^2 $}. This
is consistent with the results obtained in Ref. \cite{NISTYb2} and
 \cite{frenchboson} for the bosonic isotope. Inclusion of the
hyperfine splitting into the frequency difference in the denominator
of Eq.~\ref{Zeemanshift} yields an additional term in the second
order shift proportional to $m_F^2$ which is more that $10^{-6}$
times smaller than the main effect, and therefore negligible.
Notably, the fractional frequency shift due to the second order
Zeeman effect of{\small$~$5$\times 10^{-16}$ G$^{-2}$} is nearly
$10^8$ times smaller than that of the Cs \cite{SyrteCs, NISTCs}
clock transition, and more than an order of magnitude smaller than
that present in Hg$^{+}$ \cite{Bergquist}, Sr$^{+}$
\cite{NPLScience, NRC},and Yb$^{+}$ \cite{PTBYb, NPLYb} ion optical
clocks.

A Breit-Rabi like diagram is shown in Fig.~\ref{3p0zeeman}, giving
the shift of the $^1S_0$-$^3P_0$ transition frequency for different
$m_F$ sublevels (assuming $\Delta m=0$ for $\pi$ transitions), as a
function of magnetic field. The calculation is performed using an
analytical Breit-Rabi formula (Eq.~\ref{BRLike-formulae}) provided
in the appendix.  The result is indistinguishable from the
perturbative derivation in this section, even for fields as large as
$10^4$ G.
\section{The Effect of the Optical Lattice Potential}
In this section we consider the effect of the confining potential on
the energy shifts of the nuclear sublevels. In the presence of a
lattice potential of depth $U_T$, formed by a laser linearly
polarized along the axis of quantization defined by an external
magnetic field $B$, the level shift of a clock state ({\it
h}$\Delta_{g/e}$) from its bare energy is given by
\begin{equation}
{\small\begin{aligned} \Delta_{e}  =\, & -m_F(g_I+\delta g)\mu_0 B -
\kappa^{S}_{e}\frac {U_T}{E_R} -\kappa^{V}_{e}\xi
m_F\frac {U_T}{E_R}\\
&-\kappa^{T}_{e}\left(3m_F^2-F(F+1)\right) \frac {U_T}{E_R}
\\  \Delta_{g} =\, & - m_F g_I \mu_0 B- \kappa^{S}_{g}\frac {U_T}{E_R} -\kappa^{V}_{g}\xi
m_F\frac {U_T}{E_R}\\&-\kappa^{T}_{g}\left(3m_F^2-F(F+1)\right)
\frac {U_T}{E_R}.
\end{aligned}}\label{totalshifts}
\end{equation}
Here, $\kappa^S$, $\kappa^V$, and $\kappa^T$ are shift coefficients
proportional to the scalar, vector (or axial), and tensor
polarizabilities, and subscripts $e$ and $g$ refer to the excited
($^3P_0$) and ground ($^1S_0$) states respectively.  $E_R$ is the
energy of a lattice photon recoil and $U_T/E_R$ characterizes the
lattice intensity.  The vector ($\propto m_F$) and tensor ($\propto
m_F^2$) light shift terms arise solely from the nuclear structure
and depend on the orientation of the light polarization and the bias
magnetic field. The tensor shift coefficient includes a geometric
scaling factor which varies with the relative angle $\phi$ of the
laser polarization axis and the axis of quantization, as
3$\cos^2\phi-1$. The vector shift, which can be described as an
pseudo-magnetic field along the propagation axis of the trapping
laser, depends on the trapping geometry in two ways.  First, the
size of the effect is scaled by the degree of elliptical
polarization $\xi$, where $\xi=0$ ($\xi=\pm1$) represents perfect
linear (circular) polarization. Second, for the situation described
here, the effect of the vector light shift is expected to be orders
of magnitude smaller than the Zeeman effect, justifying the use of
the bias magnetic field direction as the quantization axis for all
of the $m_F$ terms in Eq.~\ref{totalshifts}. Hence the shift
coefficient depends on the relative angle between the
pseudo-magnetic and the bias magnetic fields, vanishing in the case
of orthogonal orientation \cite{Fortson2}.  A more general
description of the tensor and vector effects in alkaline-earth
systems for the case of arbitrary elliptical polarization can be
found in Ref.~\cite{Katori2}. Calculations of the scalar, vector,
and tensor shift coefficients have been performed elsewhere for Sr,
Yb, and Hg \cite{Katori1, Katori2, Fortson1, Fortson2} and will not
be discussed here. Hyperpolarizability effects ({\small$\propto
U_T^2$}) \cite{Anders1, Katori1, Katori2, Fortson1} are ignored in
Eq.~\ref{totalshifts} as they are negligible in $^{87}$Sr at the
level of $10^{-17}$ for the range of lattice intensities used in
current experiments \cite{Anders1}. The second order Zeeman term has
been omitted but is also present.

Using Eq.~\ref{totalshifts} we can write the frequency of a
$\pi$-transition ($\Delta m_F=0$) from a ground state $m_F$ as
\begin{equation}
{\small\begin{aligned}
 \nu_{\pi_{m_F}}  &=\, \nu_c -\left(\Delta\kappa^S-\Delta\kappa^TF(F+1)\right)\frac {U_T}{E_R}\\
&-\left(\Delta\kappa^Vm_F\xi
+\Delta\kappa^T3m_F^2\right) \frac {U_T}{E_R}\\
&-\delta g m_F\mu_0 B,
\end{aligned}}\label{pishifts}
\end{equation}
where the shift coefficients due to the differential
polarizabilities are represented as $\Delta\kappa$, and $\nu_c$ is
the bare clock frequency.  The basic principle of the lattice clock
technique is to tune the lattice wavelength (and hence the
polarizabilities) such that the intensity-dependent frequency shift
terms are reduced to zero. Due to the $m_F$-dependence of the third
term of Eq.~\ref{pishifts}, the Stark shifts cannot be completely
compensated for all of the sublevels simultaneously. Or
equivalently, the magic wavelength will be different depending on
the sublevel used.  The significance of this effect depends on the
magnitude of the tensor and vector terms. Fortunately, in the case
of the $^1S_0$-$^3P_0$ transition the clock states are nearly
scalar, and hence these effects are expected to be quite small.
While theoretical estimates for the polarizabilities have been made,
experimental measurements are unavailable for the vector and tensor
terms. The frequencies of $\sigma^{\pm}$ ($\Delta m_F=\pm1$)
transitions from a ground $m_F$ state are similar to the
$\pi$-transitions, given by
\begin{equation}
{\small\begin{aligned}
 \nu_{\sigma^{\pm}_{m_F}}  &=\, \nu_c
 -\left(\Delta\kappa^S-\Delta\kappa^TF(F+1)\right)\frac {U_T}{E_R}
 \\
 &-\left((\kappa^V_{e}(m_F\pm1)-\kappa^V_{g}m_F)\xi
\right) \frac
{U_T}{E_R}\\&-\left(\kappa^T_{e}3(m_F\pm1)^2-\kappa^T_{g}3m_F^2\right)
\frac {U_T}{E_R}\\&-(\pm g_I+\delta g(m_F\pm1))\mu_0 B.
\\
\end{aligned}}\label{sigmashifts}
\end{equation}
\section{Experimental Determination of Field Sensitivities}
\begin{figure}[t]
\resizebox{8.5cm}{!}{
\includegraphics[angle=0]{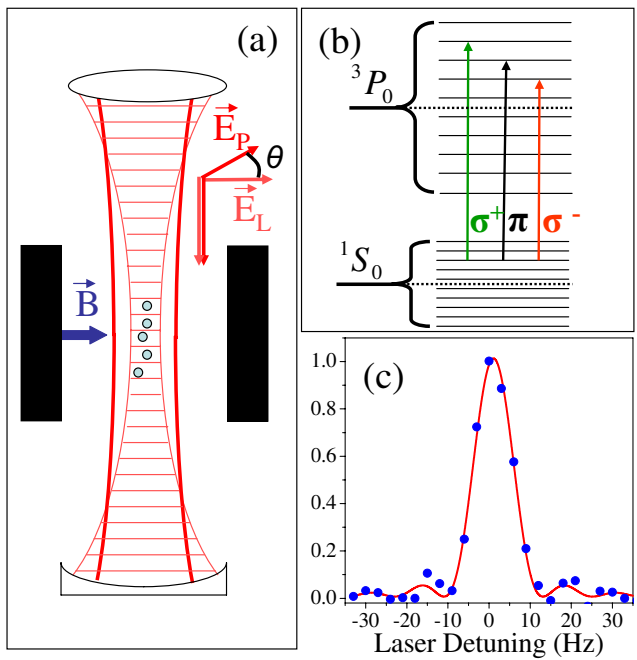}}
\caption{\label{setup}(color online) (a) Schematic of the
experimental apparatus used here. Atoms are confined in a nearly
vertical optical lattice formed by a retro-reflected 813 nm laser. A
698 nm probe laser is co-aligned with the lattice.  The probe
polarization
 $E_P$ can be varied by an angle $\theta$ relative to that of the linear lattice polarization
 $E_L$.  A pair of Helmholtz coils (blue) is used to apply a magnetic field
along the lattice polarization axis.  (b) Nuclear structure of the
$^1S_0$ and $^3P_0$ clock states.  The large nuclear spin ($I=9/2$)
results in 28 total transitions, and the labels $\pi$, $\sigma^+$,
and $\sigma^-$ represent transitions where $m_F$ changes by 0, +1,
and $-$1 respectively. (c) Observation of the clock transition
without a bias magnetic field.  The $^3P_0$ population (in arbitrary
units) is plotted (blue dots) versus the probe laser frequency for
$\theta=0$, and a fit to a sinc-squared lineshape yields a
Fourier-limited linewidth of 10.7(3) Hz. Linewidths as narrow as 5
Hz have been observed under similar conditions and when the probe
time is extended to ~500 ms. }
\end{figure}
To explore the magnitude of the various $m_F$-dependent shifts in
Eq.~\ref{pishifts},
 a differential measurement scheme can be used to eliminate the large shifts common
to all levels.  Using resolved sublevels one can extract $m_F$
sensitivities by measuring the splitting of neighboring states. This
is the approach taken here. A diagram of our spectroscopic setup is
shown in Fig.~\ref{setup}(a). $^{87}$Sr atoms are captured from a
thermal beam into a magneto-optical trap (MOT), based on the
$^1S_0$-$^1P_1$ cycling transition.  The atoms are then transferred
to a second stage MOT for narrow line cooling using a dual frequency
technique \cite{Mukaiyama1}.  Full details of the cooling and
trapping system used in this work are discussed elsewhere
 \cite{Loftus1, Ludlow1}. During the cooling process, a vertical
one-dimensional lattice is overlapped with the atom cloud. We
typically load $\sim$10$^4$ atoms into the lattice at a temperature
of $\sim$1.5$ \mu$K.  The lattice is operated at the Stark
cancelation wavelength \cite{Anders1, Boyd2} of 813.4280(5)~nm with
a trap depth of $U_0=35 E_R$.  A Helmholtz coil pair provides a
field along the lattice polarization axis for resolved sub-level
spectroscopy. Two other coil pairs are used along the other axes to
zero the orthogonal fields.  The spectroscopy sequence for the
$^1S_0$-$^3P_0$ clock transition begins with an 80 ms Rabi pulse
from a highly stabilized diode laser \cite{Ludlow2} that is
co-propagated
 with the lattice laser.  The polarization of the probe laser is linear at an angle $\theta$
 relative to that of the lattice.  A shelved detection scheme is
used, where the ground state population is measured using the
$^1S_0$-$^1P_1$ transition.  The $^3P_0$ population is then measured
by pumping the atoms through intermediate states using
$^3P_0$-$^3S_1$,$^3P_2$-$^3S_1$, and the natural decay of $^3P_1$ ,
before applying a second $^1S_0$-$^1P_1$ pulse. The 461~nm pulse is
destructive, so for each frequency step of the probe laser the
$\sim$800~ms loading and cooling cycle is repeated.

\begin{figure}[t]
\resizebox{8.5cm}{!}{
\includegraphics[angle=0]{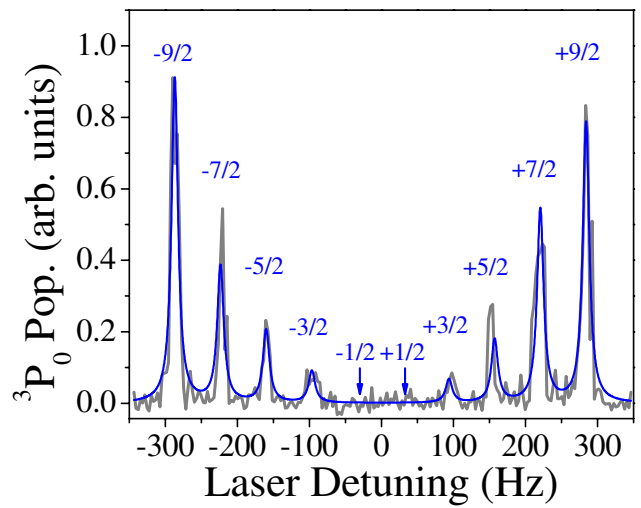}}
\caption{\label{pifig}(color online) Observation of the
$^1S_0$-$^3P_0$ $\pi$-transitions ($\theta=0$) in the presence of a
0.58 G magnetic field.  Data is shown in grey and a fit to the eight
observable lineshapes is shown as a blue curve.  The peaks are
labeled by the ground state $m_F$-sublevel of the transition.  The
relative transition amplitudes for the different sublevels are
strongly influenced by the Clebsch-Gordan coefficients.  Here,
transition linewidths of 10 Hz are used.  Spectra as narrow as 1.8
Hz can be achieved under similar conditions if the probe time is
extended to 500 ms. }
\end{figure}

When $\pi$ polarization is used for spectroscopy ($\theta=0$), the
large nuclear spin provides ten possible transitions, as shown
schematically in Fig.~\ref{setup}(b).  Figure~\ref{setup}(c) shows a
spectroscopic measurement of these states in the absence of a bias
magnetic field. The suppression of motional effects provided by the
lattice confinement allows observation of extremely narrow lines
\cite{Boyd1, NistYb, HoytEFTF}, in this case having Fourier-limited
full width at half maximum (FWHM) of $\sim$10 Hz (quality factor of
$4\times10^{13}$).  In our current apparatus the linewidth
limitation is 5 Hz with degenerate sublevels and 1.8 Hz when the
degeneracy is removed \cite{Boyd1}. The high spectral resolution
allows for the study of nuclear spin effects at small bias fields,
as the ten sublevels can easily be resolved with a few hundred mG.
An example of this is shown in Fig.~\ref{pifig}, where the ten
transitions are observed in the presence of a 0.58 G bias field.
This is important for achieving a high accuracy measurement of
$\delta g$ as the contribution from magnetic-field-induced state
mixing is negligible. To extract the desired shift coefficients we
note that for the $\pi$ transitions we have a frequency gap between
neighboring lines of
\begin{equation}
{\small\begin{aligned}
f_{\pi, m_F}= \nu_{\pi_{m_F}}-\nu_{\pi_{m_F-1}}& \\
= -\delta g \mu_0 B- \Delta\kappa^V\xi \frac {U_T}{E_R} &-\Delta\kappa^T3(2m_F-1)\frac {U_T}{E_R}. \\
\end{aligned}}\label{fpi}
\end{equation}
From Eq.~\ref{fpi}, we see that by measuring the differences in
frequency of two spectroscopic features, the three terms of interest
($\delta g$, $\Delta\kappa^V$, and $\Delta\kappa^T$) can be
determined independently. The differential $g$ factor can be
determined by varying the magnetic field. The contribution of the
last two terms can be extracted by varying the intensity of the
standing wave trap, and can be independently determined since only
the tensor shift depends on $m_F$.

While the $\pi$ transitions allow a simple determination of $\delta
g$, the measurement requires a careful calibration of the magnetic
field and a precise control of the probe laser frequency over the
$\sim $500 seconds required to produce a scan such as in
Fig.~\ref{pifig}. Any linear laser drift will appear in the form of
a smaller or larger $\delta g$, depending on the laser scan
direction. Furthermore, the measurement can not be used to determine
the sign of $\delta g$ as an opposite sign would yield an identical
spectral pattern.
\begin{figure}[t!]
\resizebox{8.5cm}{!}{
\includegraphics[angle=0]{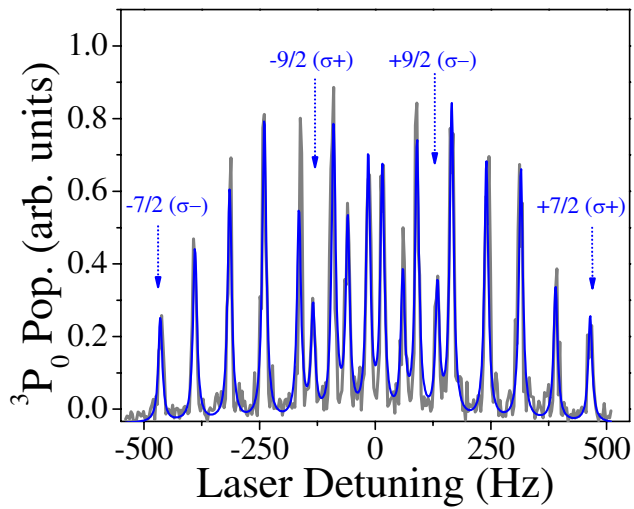}}
\caption{\label{sigfig}(color online) Observation of the 18 $\sigma$
transitions when the probe laser polarization is orthogonal to that
of the lattice ($\theta=\frac{\pi}{2}$).  Here, a field of 0.69 G is
used. The spectroscopic data is shown in grey and a fit to the data
is shown as a blue curve.  Peak labels give the ground state
sublevel of the transition, as well as the excitation polarization.}
\end{figure}
In an alternative measurement scheme, we instead polarize the probe
laser perpendicular to the lattice polarization
($\theta=\frac{\pi}{2}$) to excite both $\sigma^{+}$ and
$\sigma^{-}$ transitions.  In this configuration, 18 spectral
features are observed and easily identified (Fig.~\ref{sigfig}).
Ignoring small shifts due to the lattice potential, $\delta g$ is
given by extracting the frequency splitting between adjacent
transitions of a given polarization (all $\sigma^+$ or all
$\sigma^-$ transitions) as {\small$f_{\sigma^{\pm},
m_F}$=$\nu_{\sigma^{\pm}_{m_F}}-\nu_{\sigma^{\pm}_{m_F-1}}$=$-\delta
g \mu_0 B$ }. If we also measure the frequency difference between
$\sigma^+$ and $\sigma^-$ transitions from the same sublevel,
{\small$f_{d,
m_F}$=$\nu_{\sigma^{+}_{m_F}}-\nu_{\sigma^{-}_{m_F}}$=$-2(g_I+\delta
g)\mu_0 B$}, we find that the differential $g$-factor can be
determined from the ratio of these frequencies as
\begin{equation}
{\small\delta g =\frac{ g_I}{\frac{f_{d, m_F}}{2f_{\sigma^{\pm},
m_F}}-1}.}\label{gmeas}
\end{equation}
In this case, prior knowledge of the magnetic field is not required
for the evaluation, nor is a series of measurement at different
fields, as $\delta g$ is instead directly determined from the line
splitting and the known $^1S_0$ $g$ factor $g_I$. The field
calibration and the $\delta g$ measurement are in fact done
simultaneously, making the method immune to some systematics which
could mimic a false field, such as linear laser drift during a
spectroscopic scan or slow magnetic field variations. Using the
$\sigma$ transitions also eliminates the sign ambiguity which
persists when using the $\pi$ transitions for measuring $\delta g$.
While we can not extract the absolute sign, the recovered spectrum
is sensitive to the relative sign between $g_I$ and $\delta g$. This
is shown explicitly in Fig.~\ref{sigcalc} where the positions of the
transitions have been calculated in the presence of a $\sim $1~G
magnetic field. Figure~\ref{sigcalc}(a) shows the spectrum when the
signs of $g_I$ and $\delta g$ are the same while in
Fig.~\ref{sigcalc}(b) the signs are opposite.  The two plots show a
qualitative difference between the two possible cases. Comparing
Fig.~\ref{sigfig} and Fig.~\ref{sigcalc} it is obvious that the
hyperfine interaction \emph{increases} the magnitude of the $^3P_0$
$g$-factor ($\delta g$ has the same sign as $g_I$). We state this
point explicitly because of recent inconsistencies in theoretical
estimates of the relative sign of $\delta g$ and $g_I$ in the
$^{87}$Sr literature \cite{KatoriJSP,LeTargat1}.
\begin{figure}[t]
\resizebox{8.5cm}{!}{
\includegraphics[angle=0]{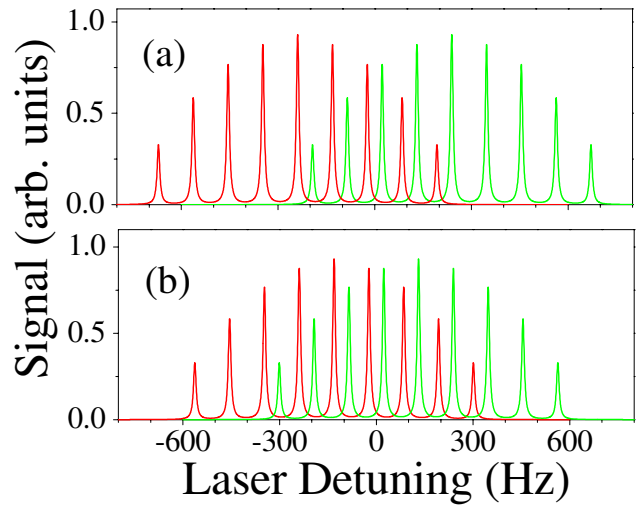}}
\caption{\label{sigcalc}(color online) Calculation of the 18
$\sigma$ transition frequencies in the presence of a 1 G bias field,
including the influence of Clebsch-Gordan coefficients.  The green
(red) curves show the $\sigma^+$ ($\sigma^-$) transitions. (a)
Spectral pattern for $g$-factors {\small$g_I\mu_0=-185$ Hz/G} and
{\small$\delta g\mu_0=-109$ Hz/G}. (b) Same pattern as in (a) but
with {\small$\delta g\mu_0=+109$ Hz/G}.  The qualitative difference
in the relative positions of the transitions allows determination of
the sign of $\delta g$ compared to that of $g_I$. }
\end{figure}

\begin{figure}[t]
\resizebox{8.5cm}{!}{
\includegraphics[angle=0]{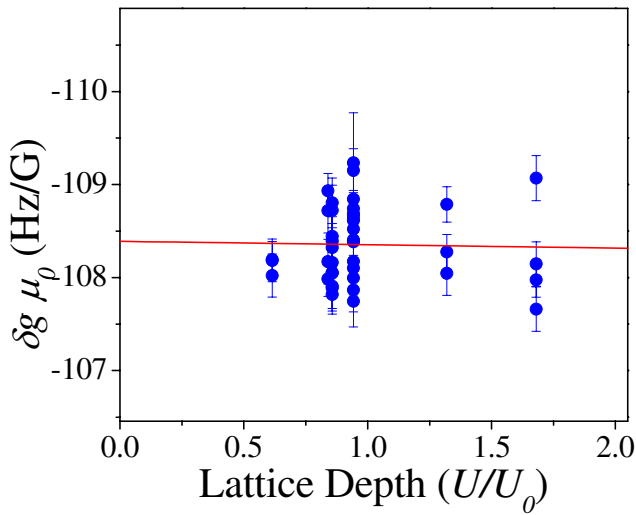}}
\caption{\label{gstark}(color online) Summary of $\delta
g$-measurements for different lattice intensities.  Each data point
(and uncertainty) represents the $\delta g$ value extracted from a
full $\sigma^{\pm}$ spectrum such as in Fig.~\ref{sigfig}. Linear
extrapolation (red line) to zero lattice intensity yields a value
$-$108.4(1) Hz/G.}
\end{figure}
To extract the magnitude of $\delta g$,  data such as in
Fig.~\ref{sigfig} are fit with eighteen Lorentzian lines, and the
relevant splitting frequencies $f_{d, m_F}$ and $f_{\sigma^{\pm}}$
are extracted. Due to the large number of spectral features, each
experimental spectrum yields 16 measurements of $\delta g$.  A total
of 31 full spectra was taken, resulting in an average value of
{\small$\delta g \mu_0=-108.4(4)$ Hz/G} where the uncertainty is the
standard deviation of the measured value. To check for sources of
systematic error, the magnetic field was varied to confirm the field
independence of the measurement.  We also varied the clock laser
intensity by an order of magnitude to check for Stark and line
pulling effects.  It is also necessary to consider potential
measurement errors due to the optical lattice since in general the
splitting frequencies $f_{d, m_F}$ and $f_{\sigma^{\pm}}$ will
depend on the vector and tensor light shifts.  For fixed fields, the
vector shift is indistinguishable from the linear Zeeman shift (see
Eqs.~\ref{pishifts}-\ref{fpi}) and can lead to errors in calibrating
the field for a $\delta g$ measurement.  In this work, a high
quality linear polarizer ($10^{-4}$) is used which would in
principle eliminate the vector shift. The nearly orthogonal
orientation should further reduce the shift. However, any
birefringence of the vacuum windows or misalignment between the
lattice polarization axis and the magnetic field axis can lead to a
non-zero value of the vector shift. To measure this effect in our
system, we varied the trapping depth over a range of {\small$\sim
(0.6-1.7) U_0$} and extrapolated $\delta g$ to zero intensity, as
shown in Fig.~\ref{gstark}. Note that this measurement also checks
for possible errors due to scalar and tensor polarizabilites as
their effects also scale linearly with the trap intensity.  We found
that the $\delta g$-measurement was affected by the lattice
potential by less then 0.1$\%$, well below the uncertainty quoted
above.

\begin{figure}[t]
\resizebox{8.5cm}{!}{
\includegraphics[angle=0]{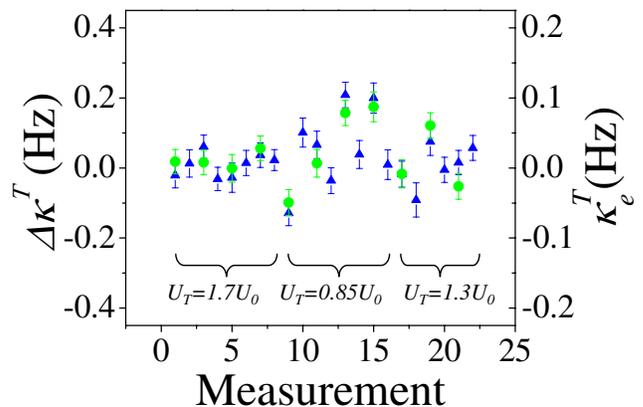}}
\caption{\label{tensorshift}(color online) Measurement of the tensor
shift coefficients $\Delta\kappa^T$ (blue triangles), and
$\kappa^T_e$(green circles), using $\sigma$ spectra and
Eq.~\ref{Tensorshiftformula}.  The measured coefficients show no
statistically significant trap depth dependence while varying the
depth from 0.85--1.7 $U_0$.}
\end{figure}
Unlike the vector shift, the tensor contribution to the sublevel
splitting is distinguishable from the magnetic contribution even for
fixed fields.   Adjacent $\sigma$ transitions can be used to measure
$\Delta\kappa^T$ and $\kappa^T_e$ due to the $m_F^2$ dependence of
the tensor shift.  An appropriate choice of transition comparisons
results in a measurement of the tensor shift without any
contributions from magnetic or vector terms.  To enhance the
sensitivity of our measurement we focus mainly on the transitions
originating from states with large $m_F$;  for example, we find that
\begin{equation}
{\small\begin{aligned} \Delta\kappa^T=&
-\frac{f_{\sigma^{+},m_F=7/2}
-f_{\sigma^{+},m_F=-7/2}}{42\frac {U_T}{E_R}}\\
\kappa^T_e=& -\frac{f_{d,m_F=7/2} -f_{d,m_F=-7/2}}{84\frac {U_T}{E_R}}, \\
\end{aligned}}\label{Tensorshiftformula}
\end{equation}

\noindent while similar combinations can be used to isolate the
differential tensor shift from the $\sigma^-$ data as well as the
tensor shift coefficient of the $^1S_0$ state. From the $\sigma$
splitting data we find {\small$\Delta\kappa^T$= 0.03(8) Hz/$U_0$}
and {\small$|\kappa^T_e|$=0.02(4) Hz/$U_0$}.  The data for these
measurements is shown in Fig.~\ref{tensorshift}.  Similarly, we
extracted the tensor shift coefficient from $\pi$ spectra,
exploiting the $m_F$-dependent term in Eq.~\ref{fpi}, yielding
{\small$\Delta\kappa^T=0.02(7)$ Hz/$U_0$}. The measurements here are
consistent with zero and were not found to depend on the trapping
depth used for a range of 0.85--1.7 $U_0$, and hence are interpreted
as conservative upper limits to the shift coefficients. The error
bars represent the standard deviation of many measurements, with the
scatter in the data due mainly to laser frequency noise and slight
under sampling of the peaks. It is worth noting that the tensor
shift of the clock transition is expected to be dominated by the
$^3P_0$ shift, and therefore, the limit on $\kappa^T_e$ can be used
as an additional estimate for the upper limit on $\Delta \kappa^T$.
 Improvements on these limits can be made by going to larger trap
intensities to enhance sensitivity, as well as by directly
stabilizing the clock laser to components of interest for improved
averaging.

Table~\ref{Measuredshifts} summarizes the measured sensitivities to
magnetic fields and the lattice potential. The Stark shift
coefficients for linear polarization at 813.4280(5)~nm are given in
units of Hz/($U_T$/$E_R$). For completeness, a recent measurement of
the second order Zeeman shift using $^{88}$Sr has been included
\cite{frenchboson}, as well as the measured shift coefficient
$\Delta\gamma$ for the hyperpolarizability \cite{Anders1} and the
upper limit for the overall linear lattice shift coefficient
$\kappa$ from our recent clock measurement \cite{Boyd2}.   While we
were able to confirm that the vector shift effect is small and
consistent with zero in our system, we do not report a limit for the
vector shift coefficient $\Delta \kappa^V$ due to uncertainty in the
lattice polarization purity and orientation relative to the
quantization axis.  In future measurements, use of circular trap
polarization can enhance the measurement precision of $\Delta
\kappa^V$ by at least two orders of magnitude.

Although only upper limits are reported here, the result can be used
to estimate accuracy and linewidth limitations for lattice clocks.
For example, in the absence of magnetic fields, the tensor shift can
cause line broadening of the transition for unpolarized samples.
Given the transition amplitudes in Fig.~\ref{pifig}, the upper limit
for line broadening, derived from the tensor shift coefficients
discussed above, is 5 Hz at $U_0$. The tensor shift also results in
a different magic wavelength for different $m_F$ sublevels, which is
constrained here to the few picometer level.
\begin{table}[t]
\caption{Measured Field Sensitivities for $^{87}$Sr}
\label{Measuredshifts}
\begin{tabular}{|c|c|c|c|}\hline
Sensitivity & Value & Units & Ref.\\
\hline
$\Delta_B^{(1)}$/$m_F B$  &-108.4(4)& Hz/G & This work \\
$\Delta_B^{(2)}$/$B^2$ & -0.233(5)& Hz/G$^2$ & \cite{frenchboson}$^a$\\
$\Delta\kappa^T$ & 6(20) $\times$10$^{-4}$& Hz/($U_T$/$E_R$)&This work$^b$\\
$\Delta\kappa^T$ & 9(23)$\times$10$^{-4}$ & Hz/($U_T$/$E_R$)& This work$^c$\\
$\kappa^T_e$ & 5(10)$\times$10$^{-4}$ & Hz/($U_T$/$E_R$)& This work$^c$ \\
 $\kappa$ & -3(7)$\times$10$^{-3}$ & Hz/($U_T$/$E_R$)& \cite{Boyd2}$^d$\\
$\Delta\gamma$ & 7(6)$\times$10$^{-6}$ & Hz/($U_T$/$E_R$)$^2$ & \cite{Anders1}$^d$\\
\hline
 \multicolumn{4}{|l|}{\small{$^a$ Measured for $^{88}$Sr}}\\
\multicolumn{4}{|l|}{\small{$^b$ Measured with $\pi $ spectra}}\\
 \multicolumn{4}{|l|}{\small{$^c$ Measured with $\sigma^{\pm}$
 spectra}}\\
\multicolumn{4}{|l|}{\small{$^d$ Measured with degenerate
sublevels}}\\
\hline
\end{tabular}
\end{table}
\section{Comparison of the $\delta g$ measurement with theory and $^3P_0$ lifetime estimate}
The precise measurement of $\delta g$ provides an opportunity to
compare various atomic hyperfine interaction theories to the
experiment.  To calculate the mixing parameters $\alpha_0$ and
$\beta_0$ (defined in Eq.~\ref{HFIdef} of the Appendix), we first
try the simplest approach using the standard Breit-Wills (BW) theory
\cite{Breit, Lurio1} to relate the mixing parameters to the measured
triplet hyperfine splitting (hfs). The parameters $\alpha$ (0.9996)
and $\beta$ ($-$0.0286(3)) are calculated from recent determinations
of the $^3P_1$ \cite{Zelevinsky} and $^1P_1$ \cite{Mickelson}
lifetimes. The relevant singlet and triplet single-electron
hyperfine coefficients are taken from Ref. \cite{Kluge1}. From this
calculation we find {\small$\alpha_0=2.37(1)\times10^{-4}$},
{\small$\beta_0=-4.12(1)\times10^{-6}$}, and
{\small$\gamma_0=4.72(1)\times10^{-6}$}, resulting in {\small$\delta
g\mu_0 =-109.1(1)$ Hz/G} . Using the mixing values in conjunction
with Eq.~\ref{3p0lifetime} we find that the $^3P_0$ lifetime is
152(2)~s. The agreement with the measured $g$-factor is excellent,
however the BW-theory is known to have problems predicting the
$^1P_1$ characteristics based on those of the triplet states. In
this case, the BW-theory framework predicts a magnetic dipole $A$
coefficient for the $^1P_1$ state of -32.7(2) MHz, whereas the
experimental value is -3.4(4) MHz \cite{Kluge1}. Since $\delta g$ is
determined mainly by the properties of the $^3P_1$ state, it is not
surprising that the theoretical and experimental values are in good
agreement. Conversely, the lifetime of the $^3P_0$ state depends
nearly equally on the $^1P_1$ and $^3P_1$ characteristics, so the
lifetime prediction deserves further investigation.
\begin{table}[b]
\caption{Theoretical estimates of $\delta g$ and $\tau^{^3P_0}$ for
$^{87}$Sr} \label{Statemixing}
\begin{tabular}{|c|c|c|c|c|c|}\hline
\multicolumn{6}{|c|}{\it{Values used in Calculation}}\\
\multicolumn{6}{|c|}{$\alpha = 0.9996$  $\beta = -0.0286(3)$ }
\\ \hline
Calc. & $\alpha_0$ & $\beta_0$ & $\tau^{\small{^3P_0}}$  & $\delta
g\mu_0$ & $A^1P_1$  \\ &\small{$\times10^{4}$}& \small{$\times10^{6}$}&(s)& $m_F$(Hz/G) & (MHz)\\
\hline  BW & 2.37(1) & -4.12(1) & 152(2) & -109.1(1) &-32.7(2)\\
MBW I & 2.56(1) & -5.5(1) & 110(1) & -117.9(5) & -3.4(4)$^a$ \\
MBW II &2.35(1) & -3.2(1) & 182(5) & -108.4(4)$^b$& -15.9(5) \\
Ref \cite{Porsev}&--- &--- & 132 & ---& ---\\
Ref \cite{Greene,Santraprivate}& 2.9(3) & -4.7(7) & 110(30) & -130(15) $^c$ & --- \\
Ref \cite{Katori1, KatoriJSP}& ---& ---& 159  & 106$^d$ &--- \\
\hline \multicolumn{6}{|l|}{\small{$^a$ Experimental value
\cite{Kluge1}}}\\
 \multicolumn{6}{|l|}{\small{$^b$ Experimental
value from this work}}\\
 \multicolumn{6}{|l|}{\small{$^c$ Calculated using
 Eq.~\ref{dgeq}}}\\
\multicolumn{6}{|l|}{\small{$^d$ Sign inferred from Figure 1 in Ref.
\cite{KatoriJSP}}}\\ \hline
\end{tabular}
\end{table}

A modified BW (MBW) theory \cite{Lurio2, Kluge1, Lahaye} was
attempted to incorporate the singlet data and eliminate such
discrepancies. In this case $^1P_1$, $^3P_1$, and $^3P_2$ hfs are
all used in the calculation, and two scaling factors are introduced
to account for differences between singlet and triplet radial
wavefunctions when determining the HFI mixing coefficients (note
that $\gamma_0$ is not affected by this modification). This method
has been shown to be successful in the case of heavier systems such
as neutral Hg \cite{Lahaye}. We find
{\small$\alpha_0=2.56(1)\times10^{-4}$} and {\small$\beta_0=-
5.5(1)\times10^{-6}$}, resulting in {\small$\delta g \mu_0=-
117.9(5)$ Hz/G} and {\small $\tau^{^3P_0} =110(1)$} s. Here, the
agreement with experiment is fair, but the uncertainties in
experimental parameters used for the theory are too small to explain
the discrepancy.

Alternatively, we note that in Eq.~\ref{dgeq}, $\delta g$ depends
strongly on $\alpha_0\alpha$ and only weakly ($<1\%$) on
$\beta_0\beta$, such that our measurement can be used to tightly
constrain {\small$\alpha_0=2.35(1)\times10^{-4}$}, and then use only
the triplet hfs data to calculate $\beta_0$ in the MBW theory
framework. In this way we find
{\small$\beta_0=-3.2(1)\times10^{-6}$}, yielding
{\small$\tau^{^3P_0} =182(5)$s}. The resulting $^1P_1$ hfs $A$
coefficient is $-$15.9(5) MHz, which is an improvement compared to
the standard BW calculation. The inability of the BW and MBW theory
to simultaneously predict the singlet and triplet properties seems
to suggest that the theory is inadequate for $^{87}$Sr.  A second
possibility is a measurement error of some of the hfs coefficients,
or the ground state $g$-factor. The triplet hfs is well resolved and
has been confirmed with high accuracy in a number of measurements.
An error in the ground state $g$-factor measurement at the 10$\%$
level is unlikely, but it can be tested in future measurements by
calibrating the field in an independent way so that both $g_I$ and
$\delta g$ can be measured. On the other hand, the $^1P_1$ hfs
measurement has only been performed once using level crossing
techniques, and is complicated by the fact that the structure is not
resolved, and that the $^{88}$Sr transition dominates the spectrum
for naturally abundant samples. Present $^{87}$Sr cooling
experiments could be used to provide an improved measurement of the
$^1P_1$ data to check whether this is the origin of the discrepancy.

Although one can presumably predict the lifetime with a few percent
accuracy (based on uncertainties in the experimental data), the
large model-dependent spread in values introduces significant
additional uncertainty. Based on the calculations above (and many
other similar ones) and our experimental data, the predicted
lifetime is 145(40)~s. A direct measurement of the natural lifetime
would be ideal, as has been done in similar studies with trapped ion
systems such as In$^+$ \cite{Becker} and Al$^+$ \cite{Till} or
neutral atoms where the lifetime is shorter, but for Sr this type of
experiment is difficult due to trap lifetime limitations, and the
measurement accuracy would be limited by blackbody quenching of the
$^3P_0$ state \cite{Xu}.

Table~\ref{Statemixing} summarizes the calculations of $\delta g$
and $\tau^{^3P_0}$ discussed here including the HFI mixing
parameters $\alpha_0$ and $\beta_0$. Other recent calculations based
on the BW theory \cite{Katori1, KatoriJSP}, {\it ab initio}
relativistic many body calculations \cite{Porsev}, and an effective
core calculation \cite{Greene} are given for comparison, with error
bars shown when available.

\section{Implications for the $^{87}$Sr lattice clock}
\begin{figure}[t]
\resizebox{8.5cm}{!}{
\includegraphics[angle=0]{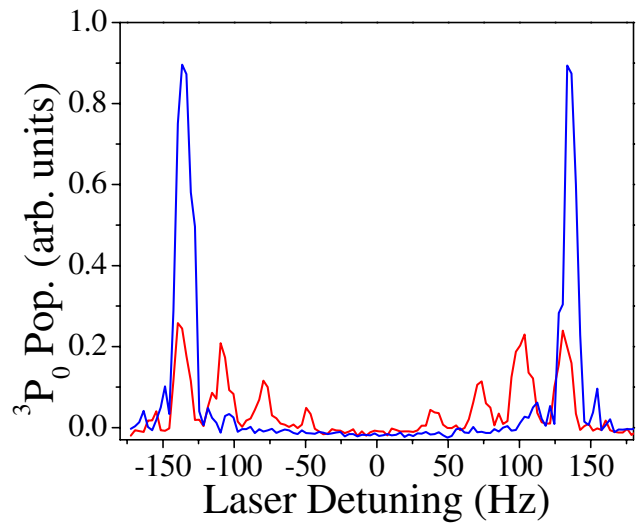}}
\caption{\label{Polarizing}(color online) The effect of optical
pumping via the $^3P_1$ ($F=7/2$) state is shown via direct
spectroscopy with $\theta =0$.  The red data shows the spectrum
without the polarizing light for a field of 0.27 G.  With the
polarizing step added to the spectroscopy sequence the blue spectum
is observed. Even with the loss of $\sim 15 \%$ of the total atom
number due to the polarizing laser, the signal size of the $m_F=\pm
9/2$ states is increased by more than a factor of 4.}
\end{figure}

In the previous sections, the magnitude of relevant magnetic and
Stark shifts has been discussed.  Briefly, we will discuss
straightforward methods to reduce or eliminate the effects of the
field sensitivities. To eliminate linear Zeeman and vector light
shifts the obvious path is to use resolved sublevels and average out
the effects by alternating between measurements of levels with the
same $|m_F|$. Figure~\ref{Polarizing} shows an example of a
spin-polarized measurement using the $m_F=\pm 9/2$ states for
cancelation of the Zeeman and vector shifts. To polarize the sample,
we optically pump the atoms using a weak beam resonant with the
$^1S_0$-$^3P_1$ ($F=7/2$) transition. The beam is co-aligned with
the lattice and clock laser and linearly polarized along the lattice
polarization axis ($\theta =0$), resulting in optical pumping to the
stretched ($m_F =9/2$) states. Spectroscopy with (blue) and without
(red) the polarizing step shows the efficiency of the optical
pumping as the population in the stretched states is dramatically
increased while excitations from other sublevels are not visible.
Alternate schemes have been demonstrated elsewhere \cite{Mukaiyama1,
KatoriJSP} where the population is pumped into a single $m_F=\pm9/2$
state using the $^1S_0$-$^3P_1$ ($F=9/2$) transition. In our system,
we have found the method shown here to be more efficient in terms of
atom number in the final state and state purity. The highly
efficient optical pumping and high spectral resolution should allow
clock operation with a bias field of less than 300 mG for a 10 Hz
feature while keeping line pulling effects due to the presence of
the other sublevels below $10^{-17}$. The corresponding second order
Zeeman shift for such a field is only $\sim$21 mHz, and hence
knowledge of the magnetic field at the 10$\%$ level is sufficient to
control the effect below 10$^{-17}$. With the high accuracy $\delta
g$-measurement reported here, real time magnetic field calibration
at the level of a few percent is trivial.  For spin-polarized
samples, a new magic wavelength can be determined for the
$m_F$-pair, and the effect of the tensor shift will only be to
modify the cancelation wavelength by at most a few picometers if a
different set of sublevels are employed. With spin-polarized
samples, the sensitivity to both magnetic and optical fields
(including hyperpolarizability effects) should not prevent the clock
accuracy from reaching below $10^{-17}$.

Initial concerns that nuclear spin effects would limit the
obtainable accuracy of a lattice clock have prompted a number of
recent proposals to use bosonic isotopes in combination with
external field induced state mixing
\cite{Hong,Santra,NISTYb2,Thom,Katorimix} to replace the mixing
provided naturally by the nuclear spin.  In these schemes, however,
the simplicity of a hyperfine-free system comes at the cost of
additional accuracy concerns as the mixing fields also shift the
clock states.  The magnitudes of the shifts depend on the species,
mixing mechanism, and achievable spectral resolution in a given
system.  As an example, we discuss the magnetic field induced mixing
scheme \cite{NISTYb2} which was the first to be experimentally
demonstrated for Yb \cite{NistYb} and Sr \cite{frenchboson}.  For a
10 Hz $^{88}$Sr resonance (i.e. the linewidth used in this work),
the required magnetic and optical fields (set to minimize the total
frequency shift) result in a second order Zeeman shift of $-$30 Hz
and an ac Stark shift from the probe laser of $-$36 Hz. For the same
transition width, using spin-polarized $^{87}$Sr, the second order
Zeeman shift is less than $-$20 mHz for the situation in
Fig.~\ref{Polarizing}, and the ac Stark shift is less than 1 mHz.
Although the nuclear-spin-induced case requires a short
spin-polarizing stage and averaging between two sublevels, this is
preferable to the bosonic isotope, where the mixing fields must be
calibrated and monitored at the 10$^{-5}$ level to reach below
10$^{-17}$.  Other practical concerns may make the external mixing
schemes favorable, if for example isotopes with nuclear spin are not
readily available for the species of interest. In a lattice clock
with atom-shot noise limited performance, the stability could be
improved, at the cost of accuracy, by switching to a bosonic isotope
with larger natural abundance.

In conclusion we have presented a detailed experimental and
theoretical study of the nuclear spin effects in optical lattice
clocks. A perturbative approach for describing the state mixing and
magnetic sensitivity of the clock states was given for a general
alkaline-earth(-like) system, with $^{87}$Sr used as an example.
Relevant Stark shifts from the optical lattice were also discussed.
We described in detail our sign-sensitive measurement of the
differential $g$-factor of the $^1S_0$-$^3P_0$ clock transition in
$^{87}$Sr, yielding {\small$\mu_0 \delta g=-108.4(4)m_F$ Hz/G}, as
well as upper limit for the differential and exited state tensor
shift coefficients $\Delta\kappa^T=0.02$ Hz/($U_T$/$E_R$) and
$\kappa^T_e =0.01$ Hz/($U_T$/$E_R$). We have demonstrated a
polarizing scheme which should allow control of the nuclear spin
related effects in the $^{87}$Sr lattice clock to well below
10$^{-17}$.

We thank T. Ido for help during the early stages of the $g$-factor
measurement, and G. K. Campbell and A. Pe'er for careful reading of
the manuscript. This work was supported by ONR, NIST, and NSF.
Andrew Ludlow acknowledges support from NSF-IGERT through the OSEP
program at the University of Colorado.

\section{Appendix}
The appendix is organized as follows, in the first section we
briefly describe calculation of the mixing coefficients needed to
estimate the effects discussed in the main text.  We also include
relevant Zeeman matrix elements.  In the second section we describe
a perturbative treatment of the magnetic field on the
hyperfine-mixed $^3P_0$ state, resulting in a Breit-Rabi like
formula for the clock transition.  In the final section we solve the
more general case and treat the magnetic field and hyperfine
interaction simultaneously, which is necessary to calculate the
sensitivity of the $^1P_1$, $^3P_1$ and $^3P_2$ states.
\subsection{State mixing coefficients and Zeeman elements}
\begin{table*}
\caption{Zeeman Matrix Elements for Pure ($^{2S+1}L_J^0$) States}
\label{Zeeman}
\begin{tabular}{|l|}\hline
\textbf{Relevant Elements for the $^3P_0$ State:}\\
\hline
$\langle^{3}P_{0}^{0},F=I|H_{Z}|^{3}P_{0}^{0}, F=I\rangle$$=-g_{I}m_F\mu_{0}B$\\
$\langle^{3}P_{0}^{0},F=I|H_{Z}|^{3}P_{1}^{0},F'=I\rangle=$$(g_{s}-g_{l})m_{F}\mu_{0}B\sqrt{\frac{2}{3I(I+1)}}$\\
$\langle^{3}P_{0}^{0},F=I|H_{Z}|^{3}P_{1}^{0},F'=I+1\rangle=$$(g_{s}-g_{l})\mu_{0}B\sqrt{\frac{((I+1)^{2}-m_{F}^{2})(4I+6)}{3(I+1)(4(I^{2}+1)-1)}}$\\
$\langle^{3}P_{0}^{0},F=I|H_{Z}|^{3}P_{1}^{0},F'=I-1\rangle=$$(g_{s}-g_{l})\mu_{0}B\sqrt{\frac{(I^{2}-m_{F}^{2})(4I-2)}{3I(4I^{2}-1)}}$\\
\hline
\textbf{Relevant Diagonal Elements within $^3P_1$ Manifold:}\\
\hline
$\langle^{3}P_{1}^{0},F=I|H_{Z}|^{3}P_{1}^{0},F=I\rangle$$=\left(\frac{g_l+g_s-g_I(2I(I+1)-2)}{2I(I+1)}\right)m_F\mu_0B$\\
$\langle^{3}P_{1}^{0},F=I+1|H_{Z}|^{3}P_{1}^{0},F=I+1\rangle$$=\left(\frac{g_l+g_s-2g_I I}{2(I+1)}\right)m_F\mu_0 B$\\
$\langle^{3}P_{1}^{0},F=I-1|H_{Z}|^{3}P_{1}^{0},F=I-1\rangle$$=\left(-\frac{g_l+g_s+2g_I(I+1)}{2I}\right)m_F\mu_0
B$\\
\hline
\textbf{Relevant Diagonal Elements within $^1P_1$ Manifold:}\\
\hline
$\langle^{1}P_{1}^{0},F=I|H_{Z}|^{1}P_{1}^{0},F=I\rangle$$=\left(\frac{g_l-g_I(I(I+1)-1)}{I(I+1)}\right)m_F\mu_0
B$\\
$\langle^{1}P_{1}^{0},F=I+1|H_{Z}|^{1}P_{1}^{0},F=I+1\rangle$$=\left(\frac{g_l-g_I I}{(I+1)}\right)m_F\mu_0 B$\\
$\langle^{1}P_{1}^{0},F=I-1|H_{Z}|^{1}P_{1}^{0},F=I-1\rangle$$=\left(-\frac{g_l+g_I(I+1)}{I}\right)m_F\mu_0
B$\\
\hline
\end{tabular}
\end{table*}
The intermediate coupling coefficients $\alpha$ and $\beta$ are
typically calculated from measured lifetimes and transition
frequencies of the $^1P_1$ and $^3P_1$ states and a normalization
constraint, resulting in
\begin{equation}
{\small\begin{aligned} \frac{\alpha^2}{\beta^2}=
\frac{\tau^{^3P_1}}{\tau^{^1P_1}}\left(\frac{\nu_{^3P_1}}{\nu_{^1P_1}}\right)^3,\,
\alpha^2  +\beta^2 = 1.
\end{aligned}}\label{alphabeta}
\end{equation}
The HFI mixing coefficients $\alpha_0$, $\beta_0$, and $\gamma_0$
are due to the interaction between the pure $^3P_0$ state and the
spin-orbit mixed states in Eq.~\ref{LSstate} having the same total
angular momentum $F$.  They are defined as
\begin{equation}
{\small\begin{aligned}
\alpha_0=&\frac{\langle^3P_1,F=I|H_A|^3P_0^{0},F=I\rangle}{\nu_{^3P_0}-\nu_{^3P_1}}\\
\beta_0=&\frac{\langle^1P_1,F=I|H_A|^3P_0^{0},F=I\rangle}{\nu_{^3P_0}-\nu_{^1P_1}}\\
\gamma_0=&\frac{\langle^3P_2,F=I|H_Q|^3P_0^{0},F=I\rangle}{\nu_{^3P_0}-\nu_{^3P_2}}.\\
\end{aligned}}\label{HFIdef}
\end{equation}
Where $H_A$ and $H_Q$ are the magnetic dipole and electric
quadrupole contributions of the hyperfine Hamiltonian.  A standard
technique for calculating the matrix elements is to relate unknown
radial contributions of the wavefunctions to the measured hyperfine
magnetic dipole ($A$) and electric quadrupole ($Q$) coefficients.
Calculation of the matrix elements using BW theory \cite{Breit,
Lurio1, Kluge1, Lahaye, Becker} can be performed using the measured
hyperfine splitting of the triplet state along with matrix elements
provided in \cite{Lurio1}. Inclusion of the $^1P_1$ data (and an
accurate prediction of $\beta_0$) requires a modified BW theory
\cite{Lurio2,Kluge1,Lahaye} where the relation between the measured
hyperfine splitting and the radial components is more complex but
manageable if the splitting data for all of the states in the $nsnp$
manifold are available.  A thorough discussion of the two theories
is provided in Refs. \cite{Kluge1, Lahaye}.

Zeeman matrix elements for singlet and triplet states in the $nsnp$
configuration have been calculated in Ref. \cite{Lurio1}.
Table~\ref{Zeeman} summarizes those elements relevant to the work
here, where the results have been simplified by using the electronic
quantum numbers for the alkaline-earth case, but leaving the nuclear
spin quantum number general for simple application to different
species. Note that the results include the application of our sign
convention in Eq.~\ref{Hzeeman} which differs from that in Ref.
\cite{Lurio1}.

\subsection{Magnetic field as a perturbation}
To determine the magnetic sensitivity of the $^{3}P_{0}$ state due
to the hyperfine interaction with the $^{3}P_{1}$ and $^{1}P_{1}$
states, we first use a perturbative approach to add the Zeeman
interaction as a correction to the $|^{3}P_{0}\rangle$ state in
Eq.~\ref{3p0LSstate}.  The resulting matrix elements depend on
spin-orbit and hyperfine mixing coefficients $\alpha$, $\beta$,
$\alpha_{0}$, $\beta_0$, and $\gamma_0$.  For the $^3P_0$ state,
diagonal elements to first order in $\alpha_0$ and $\beta_0$ are
relevant, while for $^1P_1$ and $^3P_1$, the contribution of the
hyperfine mixing to the diagonal elements can be ignored.  All
off-diagonal terms of order $\beta^{2}$, $\alpha_{0}\alpha$,
$\alpha_{0}\beta$, $\alpha_{0}^{2}$, and smaller can be neglected.
Due to the selection rules for pure (LS) states, the only
contributions of the $^3P_2$ hyperfine mixing are of order
$\alpha_0\gamma_0$, $\gamma_0^2$, and $\beta_0\gamma_0$. Thus the
state can be ignored and the Zeeman interaction matrix $M_{z}$
between atomic $P$ states can be described in the
$\left\{|^{1}P_{1},F,m_{F}\rangle,|^{3}P_{0},F,m_{F}\rangle,|^{3}P_{1},F,m_{F}\rangle\right\}$
basis as
\begin{equation}
M_{z}=\left(\begin{array}{ccc}
\begin{split}\nu_{^{1}P_{1}}\end{split}
&M^{^{3}P_{0}}_{^{1}P_{1}}&
\begin{array}{cc}
0
\end{array}\\
M^{^{1}P_{1}}_{^{3}P_{0}}&\begin{array}{cc}\begin{split}&\\&\nu_{^{3}P_{0}}\\\\\end{split}\end{array}&M^{^{3}P_{1}}_{^{3}P_{0}}\\
\begin{array}{cc}
0\end{array}&M^{^{3}P_{0}}_{^{3}P_{1}}&\begin{split}&\nu_{^{3}P_{1}}\end{split}
\end{array}\right),\label{general-matrix}
\end{equation}
where we define diagonal elements as
\begin{equation}
\small{
\begin{split}
\nu_{^{3}P_{0}}=\nu_{^{3}P_{0}}^{0}+&\langle^{3}P_{0}^{0}|H_{Z}|^{3}P_{0}^{0}\rangle\\
&+2(\alpha\alpha_{0}-\beta\beta_{0})\langle^{3}P_{1}^{0},F=I|H_{Z}|^{3}P_{0}^{0}\rangle\\
\nu_{^{3}P_{1}}=\nu_{^{3}P_{1}}^{0}+&\sum_{F'}\left(\alpha^{2}\langle^{3}P_{1}^{0},F'|H_{Z}|^{3}P_{1}^{0},F'\rangle\right.\\
&\left.+\beta^{2}\langle^{1}P_{1}^{0},F'|H_{Z}|^{1}P_{1}^{0},F'\rangle\right)\\
\nu_{^{1}P_{1}}=\nu_{^{1}P_{1}}^{0}+&\sum_{F'}\left(\alpha^{2}\langle^{1}P_{1}^{0},F'|H_{Z}|^{1}P_{1}^{0},F'\rangle\right.\\
&\left.+\beta^{2}\langle^{3}P_{1}^{0},F'|H_{Z}|^{3}P_{1}^{0},F'\rangle\right).
\end{split}}
\end{equation}
 Off diagonal elements are given by
\begin{equation}
\small{
\begin{aligned}
M^{^{3}P_{1}}_{^{3}P_{0}}=M^{^{3}P_{0}}_{^{3}P_{1}}=&\alpha\sqrt{\sum_{F'}|\langle^{3}P_{1}^{0},F'|H_{Z}|^{3}P_{0}^{0},F\rangle|^{2}}\\
M^{^{1}P_{1}}_{^{3}P_{0}}=M^{^{3}P_{0}}_{^{1}P_{1}}=&\beta\sqrt{\sum_{F'}|\langle^{3}P_{0}^{0},F|H_{Z}|^{3}P_{1}^{0},F'\rangle|^{2}}.\\
\end{aligned}}
\end{equation}
The eigenvalues of Eq.~\ref{general-matrix} can be written
analytically as three distinct cubic roots
\begin{equation}
\small{
\begin{split}
\nu^{\pm}_{m_{F}}=&\frac{\nu_{0}}{3}\mp\sqrt{\nu_{0}^{2}+3\nu_{1}^{2}}\,\times\\
&
\cos\left[\frac{1}{3}\arccos\left[\mp\frac{2\nu_{0}^{3}+9\nu_{0}\nu_{1}^{2}+27\nu_{2}^{3}}{2(\nu_{0}^{2}+3\nu_{1}^{2})^{3/2}}\right]\pm\frac{2\pi}{3}\right]\\
\nu_{m_{F}}\equiv&\nu_{^{3}P_{0},m_{F}}=\frac{\nu_{0}}{3}+\sqrt{\nu_{0}^{2}+3\nu_{1}^{2}}\,\times\\
&
\cos\left[\frac{1}{3}\arccos\left[\frac{2\nu_{0}^{3}+9\nu_{0}\nu_{1}^{2}+27\nu_{2}^{3}}{2(\nu_{0}^{2}+3\nu_{1}^{2})^{3/2}}\right]+\frac{2\pi}{3}\right],
\end{split}}\label{lurio-expression}
\end{equation}
where we have
\begin{equation}
\small{
\begin{split}
\nu_{0}=&\nu_{^{3}P_{0}}+\nu_{^{3}P_{1}}+\nu_{^{1}P_{1}}\\
\nu_{1}=&\left[-\nu_{^{3}P_{0}}\nu_{^{3}P_{1}}-\nu_{^{3}P_{1}}\nu_{^{1}P_{1}}-\nu_{^{3}P_{0}}\nu_{^{1}P_{1}}+(M^{^{3}P_{1}}_{^{3}P_{0}})^{2}\right.\\
&\left.+(M^{^{1}P_{1}}_{^{3}P_{0}})^{2}\right]^{\frac{1}{2}}\\
\nu_{2}=&\left[\nu_{^{3}P_{0}}\nu_{^{3}P_{1}}\nu_{^{1}P_{1}}-\nu_{^{3}P_{1}}(M^{^{1}P_{1}}_{^{3}P_{0}})^{2}-\nu_{^{1}P_{1}}(M^{^{3}P_{1}}_{^{3}P_{0}})^{2}
\right]^{\frac{1}{3}}.
\end{split}}\label{elements1}
\end{equation}
Since the main goal is a description of the $^3P_0$ state
sensitivity, the solution can be simplified when one considers the
relative energy spacing of the three states, and that elements
having terms $\beta$, $\alpha\beta$, and smaller are negligible
compared to those proportional to only $\alpha$.  Therefore we can
ignore $M^{^1P_1}_{^3P_0}$ terms and find simplified eigenvalues
arising only from the interaction between $^{3}P_{1}$ and
$^{3}P_{0}$ that can be expressed as a Breit-Rabi like expression
for the $^3P_0$ state given by
\begin{equation}
{\small\begin{aligned}
\nu_{^3P_0,m_{F}}=&\frac{1}{2}\left(\nu_{^{3}P_{0}}+\nu_{^{3}P_{1}}\right)+\frac{1}{2}\left(\nu_{^{3}P_{0}}-\nu_{^{3}P_{1}}\right)\\
&\times\sqrt{1+4\frac{\sum_{F'}\alpha^{2}|\langle^{3}P_{0}^{0},F|H_{Z}|^{3}P_{1}^{0},F'\rangle|^{2}}{(\nu_{^{3}P_{0}}-\nu_{^{3}P_{1}})^{2}}}.
\end{aligned}}\label{BRLike-formulae}
\end{equation}
For magnetic fields where the Zeeman effect is small compared to the
fine-structure splitting, the result is identical to that from
Eq.~\ref{secondorder} of the main text.  The magnetic sensitivity of
the clock transition (plotted in Fig.~\ref{3p0zeeman})  is
determined by simply subtracting the
$\langle^{3}P_{0}^{0}|H_{Z}|^{3}P_{0}^{0}\rangle$ term which is
common to both states.
\subsection{Full treatment of the HFI and magnetic field}
\begin{figure}[t]
\resizebox{8.5cm}{!}{
\includegraphics[angle=0]{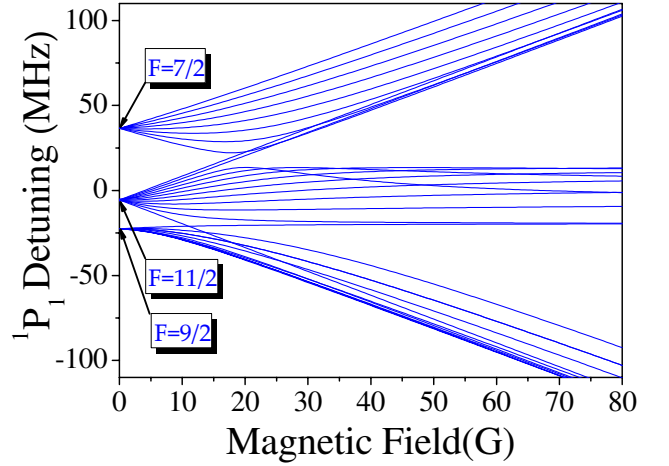}}
\caption{\label{Fig5}(color online) Magnetic sensitivity of the
$^1P_1$ state calculated with the expression in Eq.~\ref{elements2}
using $A=-3.4$ MHz and $Q=39$ MHz \cite{Kluge1}. Note the inverted
level structure.}\label{1p1}
\end{figure}
\begin{figure}[t]
\resizebox{8.5cm}{!}{
\includegraphics[angle=0]{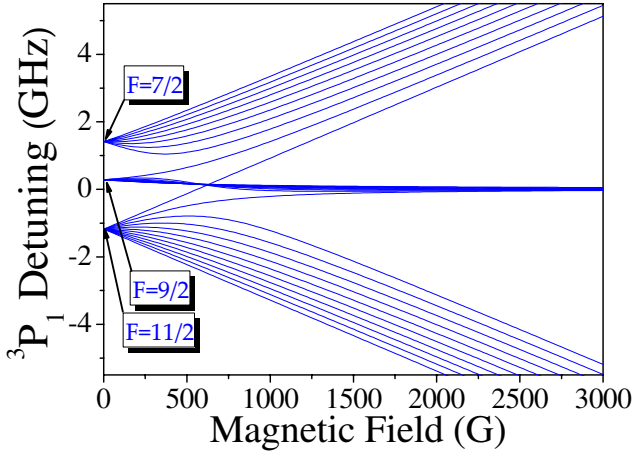}}
\caption{\label{Fig5}(color online) Magnetic sensitivity of the
$^3P_1$ state calculated with the expression in Eq.~\ref{elements2}
using $A=-260$ MHz and $Q=-35$ MHz \cite{Putlitz}.}\label{3p1}
\end{figure}
For a more complete treatment of the Zeeman effect we can relax the
constraint of small fields and treat the hyperfine and Zeeman
interactions simultaneously using the spin-orbit mixed states in
Eq.~\ref{LSstate} as a basis. The total Hamiltonian is written
$H_{\text{total}}=H_{Z}+H_{A}+H_{Q}$ including hyperfine $H_{A}$ and
quadrupole $H_{Q}$ effects in addition to the Zeeman interaction
$H_{Z}$ defined in Eq.~\ref{Hzeeman} of the main text.  The
Hamiltonian $H_{total}$ can be written as
\begin{equation}
\small{
\begin{aligned}
H_{\text{total}}=&H_Z+A\vec{I}\cdot\vec{J}\\
&+Q\frac{\frac{3}{2}\vec{I}\cdot\vec{J}(2\vec{I}\cdot\vec{J}+1)-IJ(I+1)(J+1)}{2IJ(2I-1)(2J-1)}.
\end{aligned}}\label{general-hamilonian}
\end{equation}

Diagonalization of the full space using Eq.~\ref{general-hamilonian}
does not change the $^3P_0$ result discussed above, even for fields
as large as $10^{4}$ G.  This is not surprising since the $^3P_0$
state has only one $F$ level, and is therefore only affected by the
hyperfine interaction through state mixing which was already
accounted for in the previous calculation.  Alternatively, for an
accurate description of the $^1P_1$, $^3P_1$ and $^3P_2$ states,
Eq.~\ref{general-hamilonian} must be used.  For an alkaline-earth
$^{2S+1}L_{1}$ state in the $|I,J,F,m_{F}\rangle$ basis we find an
analytical expression for the field dependence of the $F=I,I\pm1$
states and sublevels.  The solution is identical to
Eq.~\ref{lurio-expression} except we replace the frequencies in
Eq.~\ref{elements1} with those in Eq.~\ref{elements2}. We define the
relative strengths of magnetic, hyperfine, and quadrupole
interactions with respect to an effective hyperfine-quadrupole
coupling constant {\small$W_{AQ}=A+\frac{3Q}{4I(1-2I)}$} as
{\small$X_{BR}=\frac{\mu_{0}B}{W_{AQ}}$},
{\small$X_{A}=\frac{A}{W_{AQ}}$}, and
{\small$X_{Q}=\frac{Q}{I(1-2I)W_{AQ}}$}, respectively.  The solution
is a generalization of the Breit-Rabi formula \cite{Breit:1932} for
the $^{2S+1}L_{1}$ state in the two electron system with nuclear
spin $I$.  The frequencies are expanded in powers of $X_{BR}$ as
\begin{widetext}
\begin{equation}
\footnotesize{
\begin{split}
\nu_{0}&=-2W_{AQ}\left[1+\frac{3g_{I}}{2}m_{F}X_{BR}\right]\\
\nu_{1}&=W_{AQ}\sqrt{X_{\text{eff}}^{\nu_{1}}}\left[1+\frac{2(g_{\text{eff}}-g_{I})X_{A}+3g_{\text{eff}}X_{Q}}{X_{\text{eff}}^{\nu_1}}m_{F}X_{BR}+\frac{(g_{\text{eff}}+g_{I})^{2}\left(1-\frac{3m_{F}^{2}g_{I}^{2}}{(g_{\text{eff}}+g_{I})^{2}}\right)}{X_{\text{eff}}^{\nu_1}}X_{BR}^{2}\right]^{\frac{1}{2}}\\
\nu_{2}&=W_{AQ}\sqrt[3]{I(I+1)X_{\text{eff}}^{\nu_{2}}}\left[1+\frac{X_{A}^{2}\left(\frac{g_{\text{eff}}}{I(I+1)}+g_{I}\right)+X_{Q}^{2}\frac{3(1-2I)(3+2I)}{16}\left(\frac{g_{\text{eff}}}{I(I+1)}-g_{I}\right)-X_{A}X_{Q}\left(g_{\text{eff}}(2-\frac{3}{2I(I+1)})+g_{I}\right)}{X_{\text{eff}}^{\nu_2}}m_{F}X_{BR}\right.\\
&\left.\hspace{2.8cm}+\frac{m_{F}^{2}X_{A}\frac{2g_{I}g_{\text{eff}}}{I(I+1)}+X_{Q}\frac{(g_{\text{eff}}+g_{I})^{2}}{2}(1-\frac{3m_{F}^{2}g_{\text{eff}}^{2}}{I(I+1)(g_{\text{eff}}+g_{I})^{2}})}{X_{\text{eff}}^{\nu_2}}X_{BR}^{2}+\frac{\frac{g_{I}\left((g_{\text{eff}}+g_{I})^{2}-(g_{I}m_{F})^{2}\right)}{I(I+1)}}{X_{\text{eff}}^{\nu_2}}m_FX_{BR}^{3}\right]^{\frac{1}{3}},
\end{split}}\label{elements2}
\end{equation}
\end{widetext}

with abbreviations
\begin{equation}
\footnotesize{
\begin{split}
X_{\text{eff}}^{\nu_{1}}=&I(I+1)\left(X_{A}+\frac{X_{Q}}{4}-I(I+1)X_{Q}(X_{A}-1)\right)-1\\
X_{\text{eff}}^{\nu_{2}}=&X_{\text{eff}}\left(X_{Q}X_{\text{eff}}+\left(X_{A}^{2}-X_{Q}^{2}\frac{3(3+2I)(1-2I)}{16}\right)\right)\\
X_{\text{eff}}=&X_{A}+X_{Q}\frac{(3+2I)(1-2I)}{4}\\
g_{\text{eff}}=&\frac{(g_{_{l}}+g_{_{s}})}{2}+\frac{(g_{_{l}}-g_{_{s}})}{4}\left(L(L+1)-S(S+1)\right).
\end{split}}
\end{equation}
\begin{figure}[t]
\resizebox{8.5cm}{!}{
\includegraphics[angle=0]{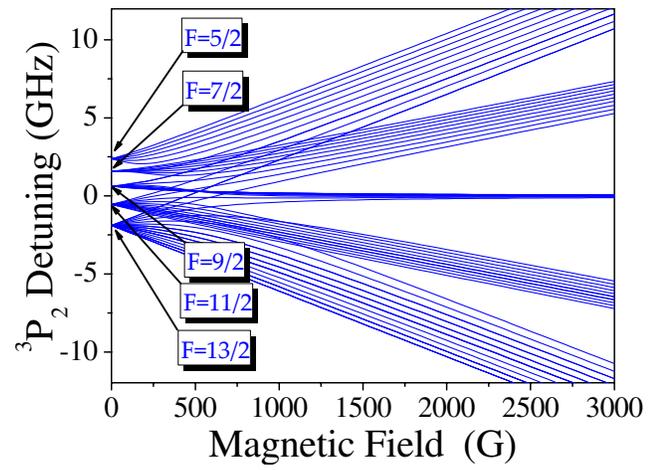}}
\caption{\label{Fig5}(color online) Magnetic sensitivity of the
$^3P_1$ state calculated numerically with
Eq.~\ref{general-hamilonian} using $A$=-212~MHz and $Q$=67~MHz
\cite{Heider:1977}.}\label{3p2}
\end{figure}

The resulting Zeeman splitting of the $5s5p^{1}P_{1}$ and
$5s5p^{3}P_{1}$ hyperfine states in $^{87}$Sr is shown in
Fig.~\ref{1p1} and Fig.~\ref{3p1}.  For the more complex structure
of $^{3}P_{2}$, we have solved Eq.~\ref{general-hamilonian}
numerically, with the results shown in
 Fig.~\ref{3p2}.  The solution for the $^1P_1$ state depends strongly on the quadrupole ($Q$) term in the Hamiltonian,
 while for the $^3P_1$ and $^3P_2$ states the magnetic dipole ($A$) term
 is dominant.

\end{document}